\documentclass[desactivate]{aa}  
\usepackage{graphicx}
\usepackage{xcolor}
\usepackage{txfonts}
\usepackage{lipsum}
\usepackage{subcaption}
\usepackage{natbib}
\usepackage[]{changes}
\usepackage{hyperref}
                               
\usepackage{lscape}

\newcommand{\adleo}{{AD~Leo}}
\newcommand{\ekdra}{{EK~Dra}}
\begin{document}

\title{Promise and pitfalls of plasma emission \\
as an exo-space-weather tool}
   \author{Ivey Davis \thanks{davis@astron.nl}}
   \institute{ASTRON, Netherlands Institute for Radio Astronomy, Oude Hoogeveensedijk 4, 7991 PD Dwingeloo, The Netherlands}
   \date{Received 24 April 2026 / Accepted 24 August 2026}

  \abstract
  {Significant observational effort has been 
  {spent on identifying} stellar analogues to solar type~II and III bursts, which are plasma emission that is produced when coronal mass ejections and fast electron beams, respectively, propagate through the corona and wind. The sensitivity of these bursts to ambient plasma and their association with energetic transient particle flux makes them powerful diagnostics of exo-solar space weather. 
  However, analyses of stellar bursts often rely on the assumption that the coronae of the stars that produce them are similar to the solar corona or that the instability requirements are the same as for the Sun. 
  By extension, the assumption is that the forms of stellar bursts are very similar to that of solar bursts.
  I introduce a simple framework and the \texttt{python} package, \texttt{swabs}, to predict the shapes of stellar type~II and III bursts to understand how they may deviate from the shapes of solar bursts. 
  I present the burst expectations in the range of 10--170\,MHz for two well-studied active stars: the M dwarf AD~Leonis, and the G dwarf EK~Draconis.
  The results for these stars suggest that the coronae of active stars may lead to bursts that have substantially (1) delayed arrival times to the observing band, (2) shallower drift rates, and (3) longer durations than solar bursts. The hot and fast winds may also inhibit the development of type~II and III bursts, regardless of the Alfv\'en speed. Non-isothermal winds reduce this inhibition, but may increase the delays and durations of the bursts.
  This work demonstrates that stellar bursts may deviate radically from their solar analogues. These deviations need to be considered in how we conduct searches for stellar bursts and how we evaluate the coronal conditions that cause the burst structures.}


\maketitle 

\section{Introduction}\label{sec:introduction}

In the thousands of exoplanet detections and many more candidates, we have identified planet hosts spanning the entire main sequence and beyond.
This broad range of spectral types leads to a similarly broad range of electromagnetic radiation environments, which change the conditions for the existence of liquid water to facilitate chemistry \citep{Kasting1993, Kopparapu2014}.
An element of these stellar environments that was often neglected but has recently received more attention is the role of the different particle fluxes in exoplanetary chemistry.
The Solar System shows that solar particle flux can strip planetary atmospheres \citep{Jakosky2015}, facilitates surface chemistry and weathering \citep{Hapke2001}, and ionizes atmospheres and distorts magnetic fields \citep{Gonzalez1994}.
Because the evidence that younger stars may generally have higher mass-loss rates, which would imply higher ram pressures \citep{Wood2005}, we might expect that the range of the habitable zone varies substantially from one star to the next in a given spectral type, depending on the particle environment.

It is understandable to exclude stellar particles from the habitability discussion.
It is difficult to identify stellar wind signals, and it often depends on the assumption that the wind resembles the solar wind in terms of the temperature, density, or speed, for instance.
A substantial amount of observing time is also required to find evidence of impulsive mass flux in the form of coronal mass ejections (CMEs) or other stellar energetic particle events because their occurrence is unpredictable.
The signatures of stellar CMEs are also difficult to distinguish from the flare process, which is usually assumed to accelerate the particles \citep{Leitzinger2024}.
It is even more difficult to conclusively determine whether these particles escape into the interplanetary medium \citep[][]{AlvaradoGomez2018}.
Plasma emission at low frequencies ($10{\mathrm{\,MHz}}\lesssim\nu\lesssim1000$\,MHz) is a method for identifying transient particle flux and constraining the properties of the ambient environment. 
The plasma emission process has received increasing attention and also some candidate detections \citep{Vedantham2020, AlvaradoGomez2020,  OFionnagain2022, Callingham2025, Konijn2025, Wang2026}.

Plasma emission is a coherent emission process that the Sun regularly produces in the form of type~II and III bursts.
The emission originates from the coronal plasma when a beam of particles excites it.
For type~II bursts, this beam is the shock of a CME or prominence, and for type~III bursts, the beam is in the form of electrons that are accelerated along field lines. 
We still do not fully understand the details of how Langmuir waves produced by the beam instability are converted into the observed radio waves \citep{Reid2014}, but a consequence of this conversion is that plasma emission occurs at the plasma frequency, $\nu_\text{p}=8.98\,n_e^{1/2}$\,kHz, where $n_e$ is the ambient plasma density in units of $\text{cm}^{-3}$.
Material that propagates outward through the corona and wind produces a distinctive negative frequency drift in dynamic spectra because lower-density material is excited at later times.
Thus, plasma emission provides evidence of bulk plasma motion that often leads to propagation into the interplanetary medium and information on the ambient environment \citep{Callingham2024}. Plasma emission is thus a potentially powerful tool for describing quasi-steady and transient exo-solar space weather.

Flares, CMEs, and plasma emission are related, and we might therefore expect that the stars that are most likely to produce plasma emission are those that show high levels of magnetic activity in the form of flares at other wavelengths.
 \citet{Mohan2024} recently reported a relation between solar X-ray flares and type~II radio luminosities.
This relation suggests that highly active stars that produce flares that are more luminous by some orders of magnitude than what we have observed from the Sun would also produce bursts that are substantially more luminous than solar type~II bursts.
Other analyses by \citet{Vedantham2020} asserted a similar expectation for stellar type~III bursts. This conclusion is apparently supported by work by \citet{SaintHilaire2013}, who demonstrated some correlation between the type~III burst distribution and the solar activity cycle. 
However, there have been few dedicated low-frequency observations of stars. 
Even fewer observations have been coordinated with other parts of the spectrum to conclude that the emission is associated with a flare.
Many of the low-frequency stellar detections whether from dedicated observations or from surveys are more readily attributed to electron cyclotron maser instability (ECMI) emission than to plasma emission {\citep{Lynch2017, Callingham2021, Feeney2021}}.
The apparent preference for ECMI may be partially attributed to a bias in the search method for stellar emission in surveys. 
However, the discussion of plasma emission generally lacks consideration of the exceptional coronal environments of these stars, what this means for the structure of bursts, and whether these stars being more magnetically active means that they are more likely to produce luminous plasma emission. 

In this paper, I address the ways in which stellar type~II and III bursts might significantly deviate from their solar forms. 
In addition to changing the shape of the burst, the extreme coronal environment might limit (or fully inhibit) the production of plasma emission. 
The limitation to the production of plasma emission presented here is distinct from the discussions by \citet{Villadsen2019} and \citet{AlvaradoGomez2020}, for example, who noted the role of the magnetic field in limiting the development of the super-Alfv\'enic shocks needed for type~II bursts. 
The limitations I lay out here do not explicitly depend on (or are at least less limited by) the Alfv\'en speed. 
I describe a simple one-dimensional coronal model in Sect.~\ref{sec:model}.
As a computationally light-weight model, it is straightforward to investigate a wide and dense parameter space related to stellar mass, radius, coronal temperature, magnetic field strength, and coronal density.
For the purposes of this paper, I explore the extreme ways in which burst shapes can deviate for two highly active stars for which constraints on the coronal and magnetic field properties exist: the M dwarf AD~Leonis (\adleo), and the G dwarf EK~Draconis (\ekdra).
 The former is a popular M dwarf for studying stellar radio phenomena \citep{Jackson1990, Osten2006, Villadsen2019, Zhang2026}, and the latter has been robustly studied as a young solar analogue \citep{Dorren1994, Gudel1994}. Both stars have candidate CME detections \citep{Houdebine1990, Namekata2025}.
I present the type~II and III burst profiles for these stars  in Sect.~\ref{sec:results}.
I also provide results for the Sun to validate that the model produces reasonable burst predictions.
I explore the significance of these results as they relate to deriving information of stellar coronae and implications for coordinated observations in Sect.~\ref{sec:discussion}.

\section{The model}\label{sec:model}
Although there has been substantial development in three-dimensional stellar wind and CME modelling \citep[e.g. ][]{Toth2012, vdHolst2014, AlvaradoGomez2018, Chen2025}, I approached this work from a simple one-dimensional perspective.
A one-dimensional approach should be sufficient for introducing the concept that winds can drastically change burst structures, and it makes the model more computationally accessible for those attempting to analyse bursts that have been observed.

\subsection{Wind modelling}
For the one-dimensional wind model, I focused on the radial evolution of the wind speed and neglected rotation. These solutions only depend on the stellar mass and radius ($M_\star$ and $R_\star$, respectively), the base coronal temperature ($T_0$), the mean molecular mass ($\mu$), and the polytropic index ($\gamma$).

\citet{Shi2022} presented the solution for polytropic indices $1<\gamma<5/3$ in their equations 20 and 21, starting with solving for the normalised critical point, $s_c = r_c/r_0$, where $r_c$ is the physical distance of the critical point, and  $r_0$  is the boundary of the isothermal layer, which is taken to be $R_\star$,
\begin{equation}\label{eq:polytrope1}
    \frac{1}{2} \left[ \left(\frac{C_g}{C_0}\right)^{\frac{4}{\gamma -1}} s_c^{\frac{2(2\gamma-3)}{\gamma -1}} -1\right] +\frac{1}{\gamma -1} \left( \frac{C_0^2}{C_g^2} -1\right) = 2(s_c-1).
\end{equation}
The value for $s_c$ can then be used to solve for velocity, $v$, as a function of normalised distance $s=r/r_0$,
\begin{equation}\label{eq:polytrope2}
    \frac{1}{2}(v^2 - v_c^2) = -\frac{v_c^2}{\gamma -1}\left[ \left( \frac{v_c^2s_c^2}{Vs^2} \right)^{\gamma-1} -1\right] + 2 C_g^2(\frac{1}{s} - \frac{1}{s_c}),
\end{equation}
where $v_c = GM_\star/2r_c$ for a spherically expanding wind, $C_g = \sqrt{G\,M_\star/2\,R_\star}$, and $C_0=\gamma\,k_\text{B}T_0/m_\text{p}\mu$, $m_\text{p}$ is the proton mass, and $k_\text{B}$ is the Boltzmann constant.
For $\gamma=1$, the profile is isothermal, and the Parker wind model described by \citet{Parker1965} is reproduced.

With the wind speed determined, the density and temperature profiles can be estimated.
To ensure that mass is conserved, I have that 
    $\rho(r)\,v(r)\,r^2 = \rho_0\,v_0\,R_\star^2 $,
where $\rho$ is the mass density, and a $0$ subscript indicates the values at the stellar surface. 
I assumed that the wind is composed of hydrogen and helium, that the wind is fully ionised, and that ions and electrons are accelerated together so that $\rho = 2n_e\,m_p/(X+1)$, where $X=(4/\mu - 3)/5 $ is the hydrogen mass fraction.
The temperature profile can then be calculated using the polytropic relation $T(r) = T_0(\rho(r)/\rho_0)^{\gamma-1}$.

I assumed that the stellar magnetic field is dipolar such that the field strength $B(r)=B_0(R_\star/r)^3$ in the closed-field regime, where $B_0$ is the surface magnetic field strength.
I assumed that the field is closed everywhere that $v<v_\text{A}$, where $v_\text{A}(r) = B(r)(4\pi\rho(r))^{-1/2}$ is the Alfv\'en speed.
Beyond the point that $v\geq v_\text{A}$, I followed \citet{Konijn2025} and assumed that the field is opened and that the field strength then evolves $\propto r^{-2}$.
Because magnetic processes almost certainly dominate coronal heating  \citep{Hollweg1990}, I assumed that the distance where the fields are opened is also where thermal acceleration of the wind ceases and that the wind speed is constant with distance.

\subsection{Burst modelling}
The shape of a burst in a dynamic spectrum depends on the frequencies that are excited, which in turn depend on the location and size of the exciting beam.
We know from solar observations that the physical extents of CMEs and electron beams grow as they propagate \citep{Reid2014, Temmer2021}. 
Similarly, the extent of the related radio sources is also larger at lower frequencies, or equivalently, at larger distances from the Sun \citep{Gopalswamy2000, SaintHilaire2013}.
With this growth in mind, when modelling the dynamic spectrum of a burst, I assumed that
\begin{itemize}
    \item the exciting beam propagates linearly through the wind, which is the equivalent of the beam travelling along the path of the Parker spiral
    \item the speed of the beam is constant
    \item the size of the beam increases linearly with distance
    \item all plasma in the beam extent is excited into producing plasma emission.
\end{itemize}  
To model the beam growth, I defined its linear size as $w = w_0(1 + f_w\,v_b\,t/h_0)$, where $w_0$ is the initial width, $h_0$ is the initial height of the beam, $v_b$ is the speed of the beam, $t$ is the time since the beam is accelerated, and $f_w$ is a constant factor that sets how fast the beam grows.
The goal of this modelling is to predict the general timing and shape of the bursts and not to evaluate the luminosity or the conditions for forming the plasma instability, for instance.
Although these assumptions greatly simplify the complex beam dynamics observed for the Sun, I demonstrate in Sect.~\ref{sec:results.solar_pred} that they replicate solar burst profiles reasonably well.

It is also worth noting that the physics of type~III burst production is understood much better than that of type~II bursts; there has been robust modelling of type~III electron beams and Langmuir wave evolution for the Sun \citep{Reid2013, Ratcliffe2014}.
The expectations for the drift rate and duration of solar type~III bursts at a given frequency are also releatively well constrained \citep{AlvarezHaddock1973, AlvarezHaddock1973DriftRate}. 
However, the evaluation of the fluid equations for beam and wave evolution is non-trivial, and the empirical relations derived in \citet{AlvarezHaddock1973} and \citet{AlvarezHaddock1973DriftRate} implicitly incorporate the solar corona conditions.
I therefore {treated} the beams for type~III bursts in the same way as type~II bursts.
The \texttt{python} code for modelling winds and bursts is available through the Stellar Wind and Burst Structure (\texttt{swabs}) package on Github.

\section{Results}\label{sec:results}
I present the results of the model for frequencies 10--170\,MHz.
This range is relevant to available low-frequency surveys such as the Low Frequency Array (LOFAR) Two-metre Sky Survey.
This survey was conducted between 120--168\,MHz over 8\,hr pointings with 12--24\,kHz and a 1--2\,s spectral and time resolution, respectively \citep{Shimwell2017}.
The lower end of the frequency range is relevant to the LOFAR Low Band Array  and Decametre sky surveys, conducted at 42 to 66\,MHz and 16 to 30\,MHz, respectively \citep{LoLSS, LoDESS}. It is also relevant to the Owens Valley Radio Observatory Long Wavelength Array (OVRO-LWA), which continuously images the entire sky from 13--87\,MHz at a 10\,s cadence with up to 24\,kHz spectral resolution.
The OVRO-LWA additionally benefits from operating alongside a dedicated optical monitor in order to characterise the flares that might cause the bursts \citep{Davis2025, Davis2026}.
Although there are reports of plasma emission at higher frequencies, the high cadence and long time baselines of the LOFAR surveys and the OVRO-LWA make them particularly well equipped for burst searches. I therefore focus on their frequencies here.

\subsection{Replicating solar quantities}\label{sec:results.solar_pred}
To confirm that the model might lead to reasonable expectations for other stars, I started by attempting to replicate the solar wind and bursts.
For the solar model, I assumed an isothermal wind, $\mu=0.6$, $T_0=2\times10^6\,$K, and $n_0=10^9\,\text{cm}^{-3}$.
Instead of using the magnetic field to estimate the distance of the Alfv\'en radius, I manually set the Alfv\'en radius to be $20\,R_\odot$. This quantity is informed by established measurements and predictions for the Sun \citep{Cranmer2023}.
The resulting wind solution has a speed beyond the Alfv\'en radius $v_\infty=426$\,km\,s$^{-1}$. This is close to the traditionally used value of 400\,km{\,s$^{-1}$} \citep{McComas2003}.
The wind speed result and quantities that are relevant for observations are reported in Table~\ref{tab:stellar_coronae}.

To model the solar bursts, I assumed that the beam originates at a height of $h_0 = 0.1\,R_\odot$ above the solar surface and that the physical size begins as $w_0 = 0.1\,R_\odot$ centred at this height.
The height is informed by the scale height of coronal loops in solar active regions, and the width is informed by the size of type~II sources at frequencies relevant to the starting height used here \citep{Aschwanden2000, Gopalswamy2000}.
I used a value of $f_w = 0.5$, which leads to a beam size of $\approx0.5\,$au at a distance of 1\,au. This beam size is within order unity of CME sizes at 1\,au \citep{Temmer2021} and similarly matches source sizes of metric type~II and III bursts \citep{Gopalswamy2000, Reid2014}.
Because driving a shock is so important for the production of type~II bursts, I {assumed} a CME speed of $1.7\,v_\infty=724\,$km\,s$^{-1}$, which is a representative value for low-frequency type-II-producing CMEs \citep{Maguire2020, Mann2022}.
For the type~III burst, I assumed an electron beam speed of $0.5\,c$, where $c$ is the speed of light \citep{Dulk1985}.

Because the properties of solar type~III bursts are relatively well constrained, I first focused on how the modelled type~III burst compares to solar expectations. 
I fitted the duration, $\Delta t$, of the burst at a given frequency $\nu$ with a function such as
\begin{equation}\label{eq:duration}
    \Delta t = 10^{a}\,(\nu/\text{Hz})^{-\alpha}\,\text{s}.
\end{equation}
The best fit from the \texttt{scipy} \texttt{curve\_fit} function \citep{Scipy2020} returns $a = 7.64\pm0.03$ and $\alpha = 0.941\pm0.004$. 
These values are exceptionally close to those used in the empirically derived relation $\Delta t = 10^{7.71}(\nu/\text{Hz})^{-0.95}\,$s from \citet{AlvarezHaddock1973}.
I similarly fitted the drift rate of the type~III burst as 
\begin{equation}\label{eq:drift_rate}
    \dot\nu = -b\nu^\beta
\end{equation}
and obtained $b=0.0082\pm0.0002$ and $\beta=1.868\pm0.004$. These values are consistent with the empirical drift rate from \citet{AlvarezHaddock1973DriftRate}, who reported $b = 0.01$ and $\beta=1.84$. This shows that
treating the type~III electron beam like a CME produces a type~III burst that is remarkably similar to bursts seen in solar observations.

There are less well-constrained relations for the type~II burst drift rate and duration as a function of frequency due to the broad range of speeds and geometries of CMEs. 
However, the duration of the type~II burst modelled here was 2.6\,min and 44.6\,min at 170\,MHz and 10\,MHz, respectively, which is generally consistent with solar observations.
Because the wind and bursts modelled here closely match solar observations, I used these models with confidence to estimate the burst profiles of other stars. 
It is worth acknowledging, however, that the density far beyond the Alfv\'en radius is overestimated by a factor of $\sim100$ in the \texttt{swabs} model.
The burst modelling might only work so well for the Sun because the excitation occurs so low in the corona, well before the Alfv\'en radius.
Although the wind model might not be able to replicate all properties of the solar wind, it appears to be a reasonable approach to implement a plasma beam that linearly grows through the wind to estimate type~II and III burst structures, at least within the Alfv\'en radius.

\subsection{EK~Dra and AD~Leo estimates}
\begin{figure}
    \centering
    \includegraphics[width=\linewidth]{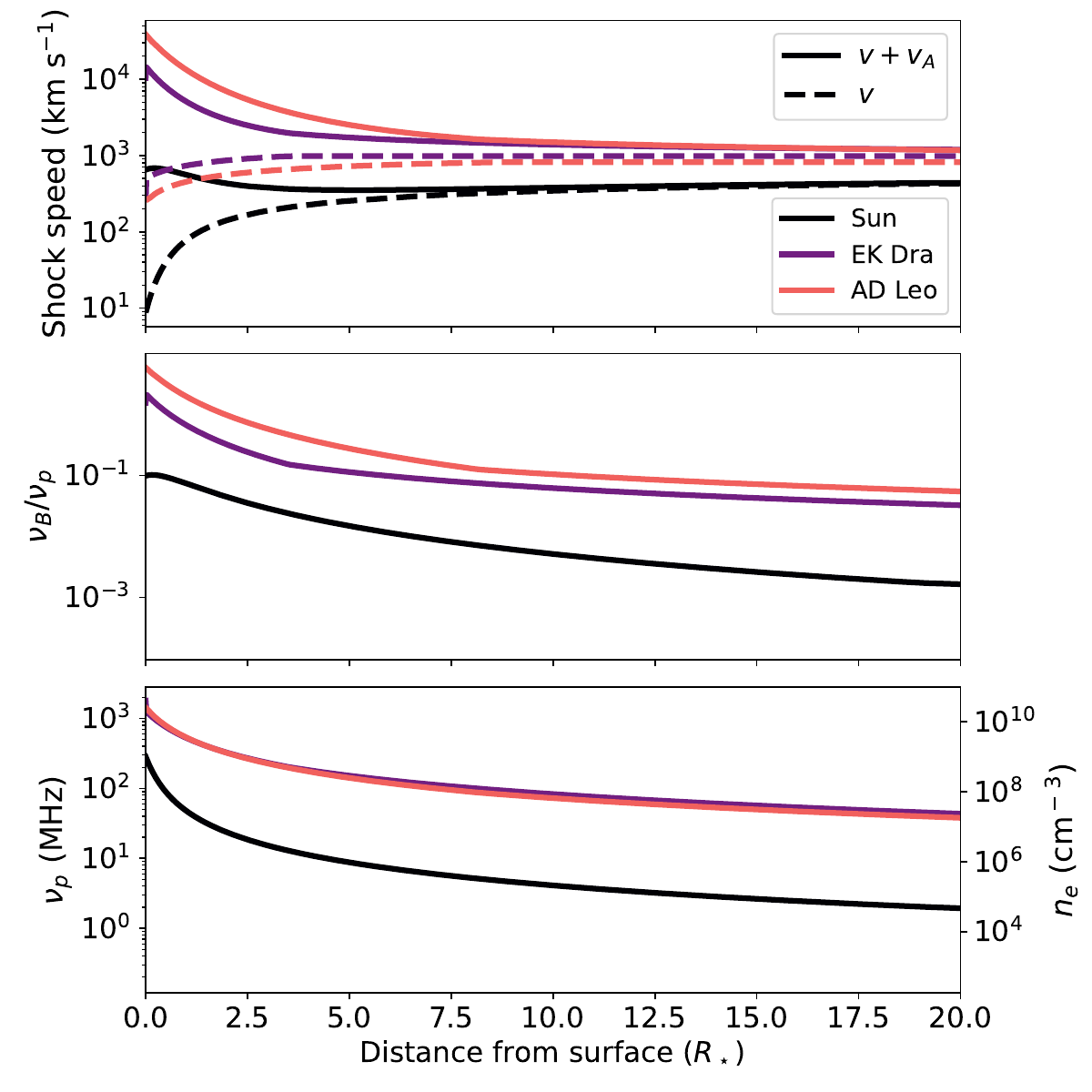}
    \caption{Wind property estimates for the Sun, EK~Dra, and AD~Leo as a function of distance from the stellar surface. The shock speed is how fast a CME must move to produce a super-Alfv\'enic shock,  and $\nu_\text{B}/\nu_\text{p}\gg1$ indicates regions where magnetic emission processes would dominate plasma emission. }
    \label{fig:general_wind_properties}
\end{figure}

To model EK~Dra and AD~Leo, I continued to follow the literature standard of an isothermal wind \citep[e.g. ][]{Ratcliffe2014, Konijn2025, Callingham2025}. 
I present the impact of $\gamma>1$  in {Section}~\ref{app:polytropic_sols}.
As with modelling the {solar wind}, I used $\mu=0.6$.
The stellar properties along with key wind results are provided in Table~\ref{tab:stellar_coronae}, and the wind properties are shown in Fig.~\ref{fig:general_wind_properties}. 

The high coronal temperatures of EK~Dra and AD~Leo lead to their winds being much faster than the solar wind. Additionally, the Alfv\'en speed contributes significantly to the shock requirements close to the stellar surface. 
The model additionally predicts shorter Alfv\'en radii than the solar Alfv\'en radius, which may be consistent with more robust models of stellar magnetospheres \citep{Reville2016, Smith2026}. 
It is worth acknowledging that \citet{Reville2016} predicted that the Alfv\'en radii of young solar-type stars are larger than the Sun's, but they underestimated the solar Alfv\'en radius.
The Alfv\'en radii estimated for EK~Dra and AD~Leo might therefore be underestimates. The effect of larger Alfv\'en radii is explored in Sect.~\ref{sec:alternative_alf}. 

Although these stars have strong magnetic fields, their coronae are also estimated to be much denser than that of the Sun. 
These modelled winds therefore suggest that the plasma frequency begins to dominate the cyclotron frequency, $\nu_\text{B}=2.8\,B/$G\,MHz, still relatively close to the stellar surface.
The frequencies where $\nu_\text{p}=\nu_\text{B}$ are 698\,MHz and 327\,MHz for EK~Dra and AD~Leo, respectively. 
When we observe below these frequencies, we might expect the dominant emission process to be plasma emission and not ECMI.

\begin{table}
\caption{Stellar properties and results for the modelled coronae{.}}             
\label{tab:stellar_coronae}      
\centering                                      
\begin{tabular}{l c c c c}          
\hline\hline                        
 {Parameter} & Unit &Sun & EK~Dra & AD~Leo   \\    
\hline                                   
    $R_\star$ & $R_\odot$ & 1.00& 0.94\tablefootmark{(1)}& {0.42}\tablefootmark{(2)}\\      
    $M_\star$ & $M_\odot$ & 1.00& 0.95\tablefootmark{(1)}& 0.42\tablefootmark{(2)}\\
    {$T_0$}&{$10^6$\,K}  & 2\tablefootmark{(3)}& 10\tablefootmark{(4)}& 6\tablefootmark{(5)}\\
    {$\log n_0$}& {cm$^{-3}$}& 9.00& 10.60\tablefootmark{(4)}& 10.40\tablefootmark{(6)}\\
    $B_0$& G & 10& 1000\tablefootmark{(7)}& 3000\tablefootmark{(8)}  \\\hline
        $r_\text{A}$& $R_\star$ & 20& 4.5& 9.2\\
    $r(\nu_B/\nu_p  < 1)$ &$R_\star$ & ... & 1.6& 2.9\\

    $h_{170}$ & $R_\star$ & 0.2 & 4.4 & 4.1\\
    {$h_{10}$} & {$R_\star$} & 4.4& 90.1& 78.8\\
    
    $v_{s,170}$& km\,s$^{-1}$ & 682& 1808& 3105\\
    $v_{s,10}$& km\,s$^{-1}$ & 353& 1031& 908\\
    $v_\infty$ & km\,s$^{-1}$ & 426& 982& 814\\
\hline
\end{tabular}

\tablefoot{$v_{s,10}$ and $v_{s,170}$ are the required speeds to drive a super-Alfv\'enic shock at the distances corresponding to a frequency of 10\,MHz and 170\,MHz, respectively. $h_{10}$ and $h_{170}$ refer to the height above the stellar surface that corresponds to a frequency of 10\,MHz and 170\,MHz, respectively. Quantities such as $T_0$, $n_0$, and $B_0$ are subject to variability over stellar cycles, and thus, they can take a variety of values in the literature. The values reported here should be sufficient for approximately representing the star and demonstrating the model response to different base properties.}
\tablebib{(1)~\citet{Waite2017}; (2)~\citet{Kossakowski2022}; (3)~\citet{Reid2013};  (4) \citet{Guuedel1995}; (5)~\citet{Gudel2003}; (6)~\citet{Ness2004}; (7)~\citet{Kochukhov2020}; (8)~\cite{Bellotti2023}}
\end{table}

As with the Sun, I  assumed for the stellar bursts that the hypothetical CME speed was $1.7\,v_\infty$ when modelling type~II bursts, that the electron beam speed was $0.5\,c$ for the type~III bursts, $h_0=0.1\,R_\star$, $w_0=0.1\,R_\star$, and $f_w=0.5$.
The assumption that the CME and beam growth follows the solar form is largely driven by our lack of knowledge of stellar CME morphology and evolution, especially in the presence of the strong fields that these stars possess.

\begin{figure}
    \centering
    \includegraphics[width=\linewidth]{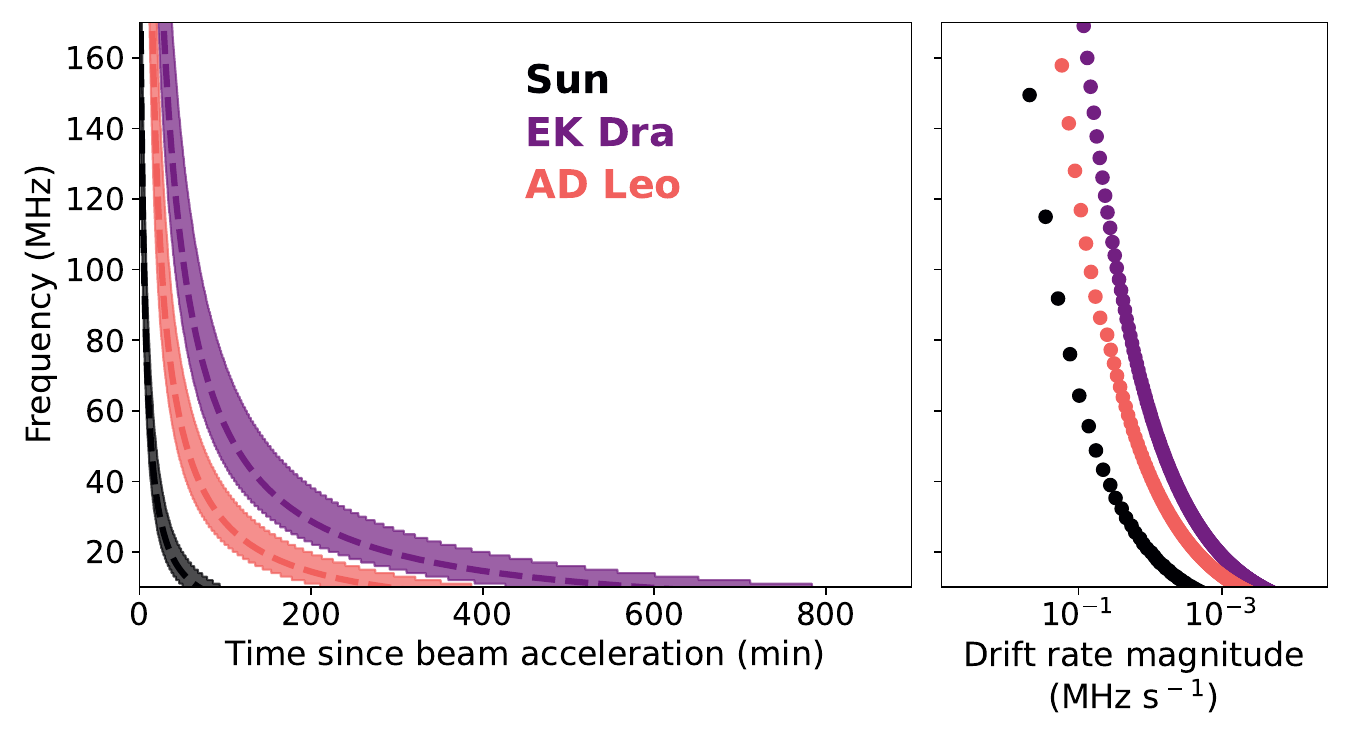}
    \caption{Type~II burst profiles for the Sun, EK~Dra, and AD~Leo (left) and the associated frequency drift rate (right). The frequency drift is estimated for the centre of the CME, with the relevant frequencies denoted by a dashed line in the left plot.  These profiles are only representative of the frequencies that would be excited and do not necessarily imply that a shock is developed. All drift rates are negative. }
    \label{fig:type_ii}
\end{figure}

\begin{figure}
    \centering
    \includegraphics[width=0.9\linewidth]{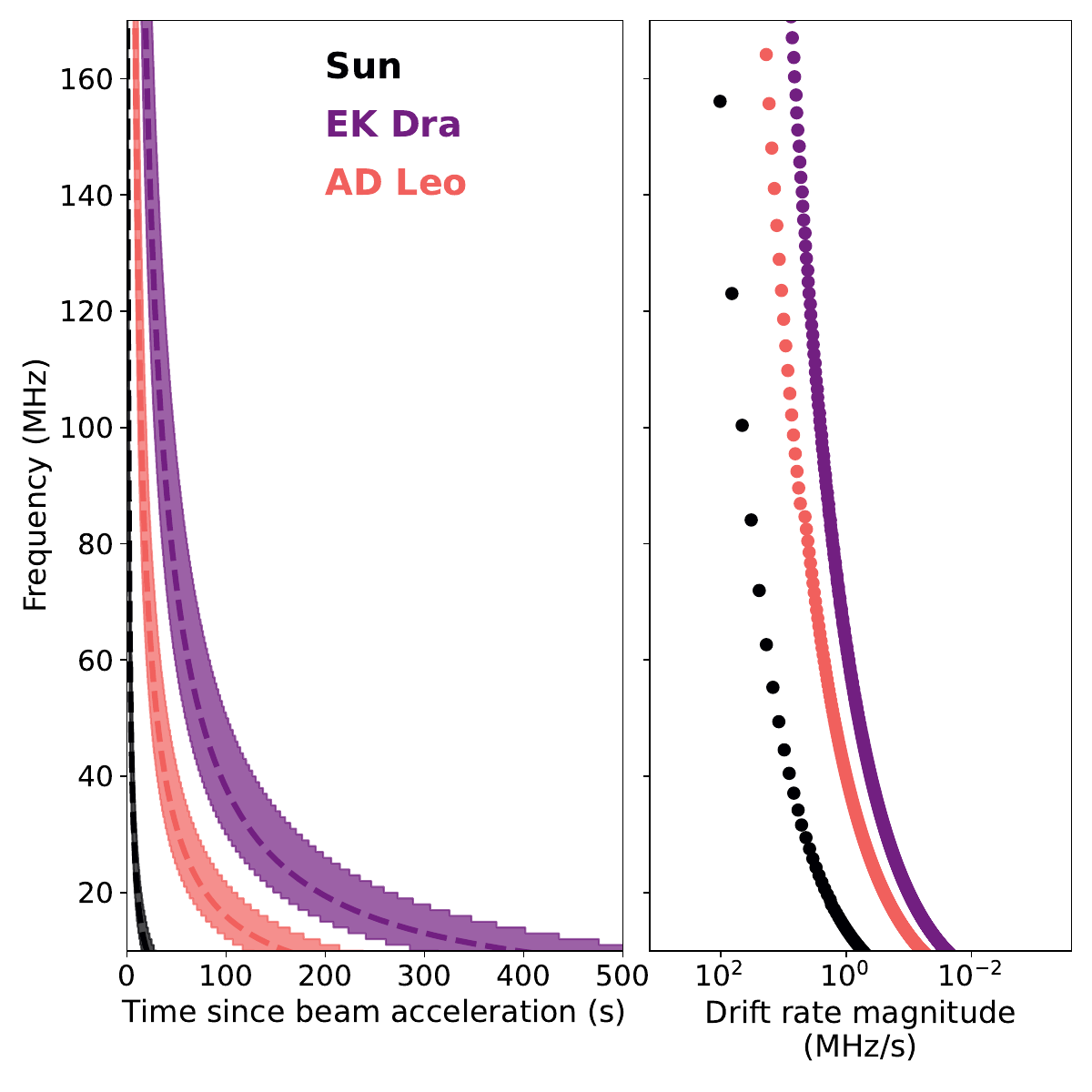}
    \caption{Same as Fig.~\ref{fig:type_ii}, but for type~III bursts.}
    \label{fig:type_iii}
\end{figure}

The resulting burst profiles and the drift rates as a function of frequency are shown in Figs.\ref{fig:type_ii} and \ref{fig:type_iii} for type~II and III bursts, respectively.
Key results for type~II and III bursts are presented in Tables~\ref{tab:type_ii} and \ref{tab:type_iii}, respectively. Perhaps the most salient result is the extreme duration and timing of the bursts relative to the beam being launched. 
This is most obvious for type~II bursts from EK~Dra, for which it may take hours to observe the burst below 100\,MHz after a CME-accelerating event such as a flare.
Similarly, the burst can last several hours at the lowest frequencies.
These results are new, but not necessarily surprising, given the properties of the stellar winds.
Namely, the extreme distances that the beams need to travel in order to reach the relevant densities and excite the associated frequencies require similarly long travel times, even when the CME moves much faster than solar CMEs.
The other key result of these models is that the magnitude of the drift rate at a given frequency is lower for active stars than it is for the Sun.
Such a substantial difference has important implications for how burst searches in dynamic spectra may be conducted, as I address in the following section.

\begin{table*}[]
    \caption{Type~II burst properties in different coronae.}
    \centering
    \begin{tabular}{lccccccc}
    \hline
    \hline
         Star & $v_\text{CME}$ & Time to $\nu_\text{170}$&Time to $\nu_\text{10}$&  $\dot\nu_\text{170}$,  &$\dot\nu_\text{10}$& $\Delta t_\text{170}$& $\Delta t_\text{10}$ \\
          & (km\,s$^{-1}$)& (min) & (min) & (MHz\,s$^{-1}$) & (MHz\,s$^{-1}$) & (min) & (min) \\
         \hline
         Sun & 724& 0.33 & 52.25 & $-0.48$&$-0.003$& 2.33 & 41.26 \\
         EK~Dra & 1\,670 & 22.00 & 447.17 & $-0.085$ & $-0.0003$ & 15.66 & 336.54\\
         AD~Leo & 1\,384 & 11.66 & 220.58 & $-0.17$ & $-0.0006$ & 8.33 & 166.27\\
        \hline
    \end{tabular}
    
    \label{tab:type_ii}
    \tablefoot{Time to frequency $\nu_i$, frequency drift rate $\dot\nu_i$, and burst duration $\Delta t_i$ were evaluated at frequencies $i = 10$, 170\,MHz.}
\end{table*}

\begin{table*}[]
    \caption{Type~III burst properties in different coronae.}
    \centering
    \begin{tabular}{lccccccc}
    \hline
    \hline
         Star & $v_\text{b}/v_e$ & Time to $\nu_\text{170}$& Time to  $\nu_\text{10}$ & $a$ & $\alpha$ & $b$ & $\beta$\\
          &  & (s)  & (s)& & & &\\
         \hline
         Sun & 19.3 & 0.1 & 15.2 & 7.64 & 0.94 & 0.01 & 1.87 \\
         EK~Dra & 8.6 & 14.7 & 299.0 & 9.60 & 1.04 & $2\times10^{-4}$ & 2.01 \\
         AD~Leo & 11.1& 6.5 & 122.3 & 9.20 & 1.04 & $5\times10^{-4}$ & 2.08 \\ 
        \hline
    \end{tabular}
    
    \label{tab:type_iii}
    \tablefoot{The time to frequency $\nu_i$ was evaluated at frequencies $i = 10$, 170\,MHz.}
\end{table*}

\section{Discussion}\label{sec:discussion}

\subsection{Implications for de-dispersion and transient searches}
When we assume that stellar type~III bursts last only as long (and drift as fast) as the solar equivalent, then current instrumentation is unlikely to be able to resolve the burst structure while also maintaining a sensitivity sufficient to detect the burst.
Instead of searching for bursts directly in dynamic spectra, de-dispersion searches can be employed \citep{Konijn2025, Davis2025}.
Similar to searches for pulsar signals, these de-dispersion searches attempt to straighten out the burst in dynamic spectra before integrating across frequency. When there is a burst, then the de-dispersion parameters that return the highest signal provide some information about the profile, even though the structure of the burst is not detected directly in the dynamic spectrum.  

\begin{figure}
    \centering
    \includegraphics[width=0.8\linewidth]{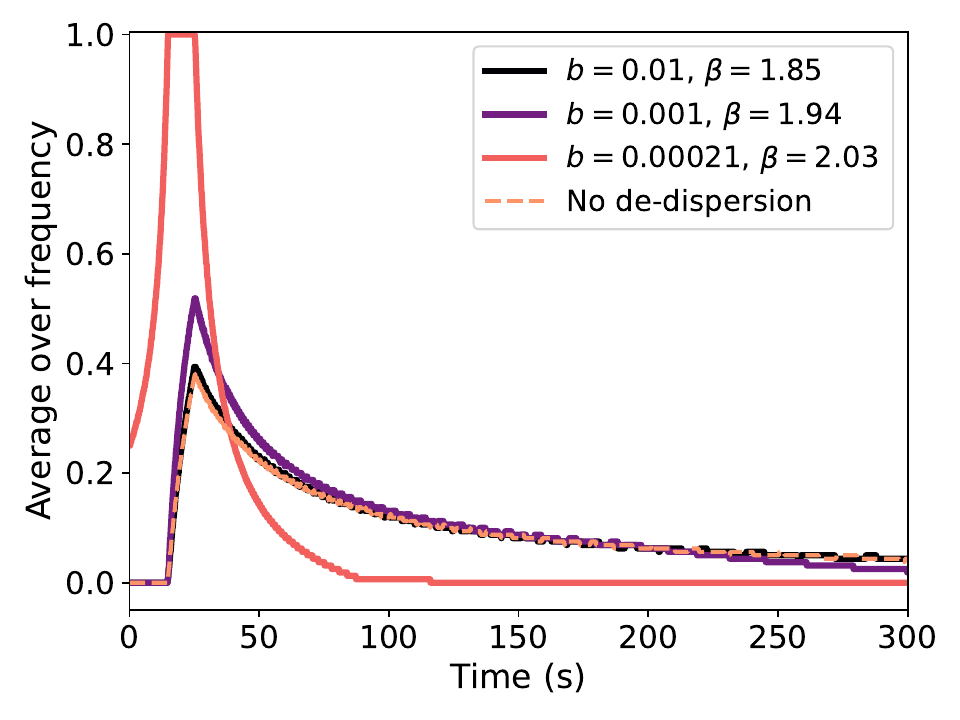}
    \caption{Frequency-averaged light curves for a type~III burst from EK~Dra when de-dispersed with various values of $b$ and $\beta$, normalised to the maximum possible value. $b = 0.01$ and $\beta=1.85$ represent empirically derived solar de-dispersion values.}
    \label{fig:dedispersion}
\end{figure}

However, Figs.~\ref{fig:type_ii} and \ref{fig:type_iii} show that the high-density coronal profiles of active stars lead to bursts that are delayed and last longer than solar bursts at the same frequencies.
For type~III bursts, a single burst can last as long as minutes at the lowest frequencies accessible from the ground, and the drift rate can be lower by an order of magnitude than what is observed for the Sun.
The effect of highly active coronae on the search for type~III bursts is thus two-fold: (1) leniency in the resolution that is required to resolve the structure, and this might allow us to observe the burst structure directly in dynamic spectra; and (2) a parameter space substantially different from solar bursts must be explored when de-dispersion searches are employed.
Related to the latter point, de-dispersion searches that only explore values near solar quantities might be missing stellar bursts.
As reported in Table~\ref{tab:type_iii}, EK~Dra and AD~Leo have $b$ values that are lower than solar values by a factor of $\approx30$--$50$. Therefore, de-dispersing with solar values might not increase the signal-to-noise ratio of a hypothetical burst.
The effect of improper de-dispersion parameters is demonstrated for EK~Dra in Fig.~\ref{fig:dedispersion}, where de-dispersing with solar values produces a peak frequency-integrated flux $\approx 0.4$ of what it would be if it were properly de-dispersed. 
The integrated flux is essentially consistent with not de-dispersing the dynamic spectrum at all when using solar values.

Because the type~II and III bursts were modelled in the same way with the exception of the beam speed, we might be able to derive and apply dispersion relations for type~II de-dispersion searches as well.
However, it is worth re-emphasising that the relative velocity distribution of CMEs, and thus the related parameter space, is likely larger than it is for type-III-producing electron beams.
Instead, we might be able to take advantage of how extreme the duration of type~II bursts could be in these coronae.
Specifically, if type~II bursts indeed last on the order of hours at a given frequency, we should not need substantial time or frequency resolution to identify such bursts.
However, a burst lasting hours might not be caught by a transient search.
There may be additional difficulties with associating such a long-duration and delayed signal with activity markers in other parts of the spectrum, as is discussed next.

\subsection{Implications for multi-wavelength coordination}\label{sec:discussion.coordination}
The increased durations and decreased drift rates of the bursts are accompanied by exceptional delays relative to the magnetic event that accelerates the particles.
Traditionally, stellar astronomers assume that this event is magnetic reconnection.
Since reconnection also produces flares, the vast majority of stellar CME candidates are accompanied by detections of flares in the optical--X-ray range.
There have been some observations of potential stellar plasma emission contemporaneous with optical flares \citep{Zic2020, Wang2026}.
However, these radio detections occurred at higher frequencies than what are considered here. These are frequencies at the threshold of where plasma emission is expected to dominate ECMI in extreme coronae, and they also correspond to distances that are still relatively low in the corona.

The perk of observing highly active stars with very dense coronae at low frequencies is that the material has to travel exceptional distances to excite the relevant frequencies. Such large distances substantially reduce the ambiguity that the material has escaped: a metric type~II burst from the Sun indicates that a coronal shock occurred, but a metric type~II burst from an active star might imply that the shock reached tens of stellar radii. 
The problem of needing to travel so far is that type~II bursts can occur several hours after the onset of a flare, complicating the multi-wavelength coordination and the means of correlating radio emission with the flare occurrence. Although simultaneous optical and low-frequency observations are crucial for testing this timing prediction, the real detection of a stellar type~II burst associated with a flare may benefit most from observations that are not simultaneous.
Although stellar type~III bursts occur later than the solar form, they might only be offset from the flare by $\lesssim10\,$min and would thus still benefit from contemporaneous observations.
Candidate type~III bursts, not type II, were indeed observed alongside optical flares by \citet{Zic2020} and \citet{Wang2026}.

It is also meaningful to consider what it means for a type~II burst to be a complement to H-$\alpha$ or other spectroscopic CME detection methods.
The models I presented here suggest that the distances from the stellar surface that are probed by  low frequencies are orders of magnitude farther than the heights probed by CME detection methods in the optical--X-ray wavelength range.
The leniency afforded by the late arrival time of the CME to low-density material offers the potential of triggered radio follow-up to flare or CME signatures.
Thus, burst properties can be used in tandem with velocity constraints from early-time CME spectroscopy to measure wind properties at distances $10\,R_\star \lesssim r\lesssim100\,R_\star$. These distances are a region of the stellar atmosphere that has been largely inaccessible to observational studies.

\subsection{Beam-speed requirements for plasma emission}
The previous discussion has focused on how the burst structure impacts observations, but it is important to reflect on the physics involved and how that might inform interpretations of candidate bursts.
In particular, the limiting factor to the production of plasma emission is the speed of the beam exciting the plasma.
This speed requirement manifests differently for type~II and III bursts and offers different means for constraining the stellar environment.

\subsubsection{Type~II shock requirements}
The speed requirement for plasma emission was introduced in the type~II model by asserting that the theoretical CME is faster than the wind speed at large distances.
However, the results provided in Table~\ref{tab:stellar_coronae} show that the actual super-Alfv\'enic speed at the distances of the relevant frequencies can make the required speed higher by a factor of a few. 
Such a high speed requirement due to strong magnetic fields was identified previously \citep{Villadsen2019, AlvaradoGomez2020}.
The speed need only be $\gtrsim1\,000\,$km{\,s$^{-1}$} for EK~Dra and AD~Leo, however, to produce a shock in the region where the plasma frequency is $\sim10$MHz. 
Such a speed is observed for some solar CMEs, and AD~Leo itself  produced a candidate CME that may have moved as fast as $\approx5\,800$\,km{\,s$^{-1}$} \citep{Houdebine1990}.
Such fast CMEs have not been reported for EK~Dra, although this might be an effect of line-of-sight derivations of the speed \citep{Namekata2025}. 

Although active stars may have fast winds, their high-density coronae soften the shock requirements imposed by high field strengths.
This softening is accomplished by decreasing the Alfv\'en speed for a given field strength and by pushing the relevant frequencies far from the stellar surface and to weaker magnetic fields.
A non-isothermal profile reduces the wind speed, further softening the shock requirements (see Sect.~\ref{app:polytropic_sols}). 
However, a non-isothermal profile also substantially changes the density profile. The distance where the density is low enough to produce emission $<170\,$MHz is twice as far for $\gamma=1.6$ as for the isothermal wind.
For a CME  meeting the minimum shock conditions, such a change to the density profile might mean that it could take several hours to more than a day for the type~II burst to be produced at the relevant frequencies.
Such a large delay between the flare and type II burst greatly complicates coordinating observations with instruments operating in other parts of the spectrum and concluding that radio emission detected with such a delay is associated with material acceleration by a flare.

\subsubsection{Type~III instability requirements}
This paper (and many others that investigated stellar type~III bursts) has so far neglected the speed requirements of the electron beam for type~III bursts.
Type~III bursts require that the exciting beam move fast enough to overcome suppression by Landau damping, requiring a minimum speed $v_\text{min} = 3\,v_e$, where $v_e = \sqrt{2\,T_e\,k_\text{B}/m_e}$ is the electron thermal velocity, $T_e$ is the electron temperature, and $m_e$ is the electron mass \citep{Reid2013}.
For the Sun, $v_\text{min}\approx0.08\,c$, and the observed lower limit of type-III-producing electron beams is indeed $\approx0.1\,c$ \citep{Dulk1985, Benz2002, Morosan2014}.
For the exceptionally hot coronae of EK~Dra and AD~Leo, $v_\text{min}=0.17$ and 0.13, respectively. 
It might then seem that a 0.5\,$c$ beam is still capable of driving the plasma instability in these coronae.
However, the growth rate of the instability is dependent on the slope of the electron velocity distribution; when the beam speed only marginally meets the speed requirement, the growth is limited and no bright burst is expected. 

Instead of focusing on the Landau damping limit, we might consider that solar electron beams have velocity contrasts $v_\text{b}/v_e\approx10$--30 and have this be the requirement.
However, there is then the issue of the beams needing to be highly relativistic.
Exceptionally fast beams can also lead to broad pitch angles \citep{Lin1981}, and the high densities would increase the collision rate and related damping \citep{Melrose1989}; both processes limit the development and saturation of the plasma instability.
Understanding whether the electron beam can facilitate the plasma instability in these extreme coronae would benefit from more rigorous modelling that addresses for instance the coupling between the energy density of Landau waves and the electrons that produce them \citep{Reid2013, Ratcliffe2014}. 
The goal here is to highlight that it is insufficient to say that any beam that moves as fast as solar beams would also drive the instability in these coronae.
Rather, physical limits of plasma emission may be used to provide constraints on the temperature in addition to the density profile of the stellar corona.

As I noted for type~II shock requirements, the wind properties change rather substantially depending on the polytropic index used, with higher indices leading to slower winds.
With regard to type~III bursts, the significant drop in temperature facilitate reaching the velocity contrasts that we see for solar type~III bursts.
While the drop in temperature and resulting change in the density profile again leads to increased delays between the time of beam acceleration and arrival of the beam to the relevant densities, the delay is likely on the order of minutes at most.
There is also a change to the relevant parameter-space scope for a de-dispersion search, but it is not nearly as drastic as the change in scope between the Sun and an active star.
Thus, a non-isothermal wind might be crucial for a star to produce a type~III burst, but  how observations or de-dispersion searches are conducted should not change substantially relative to the isothermal case.

\subsection{Alternative wind solutions}
\subsubsection{Non-isothermal wind}\label{app:polytropic_sols}
\begin{figure}
    \centering
    \includegraphics[width=\linewidth]{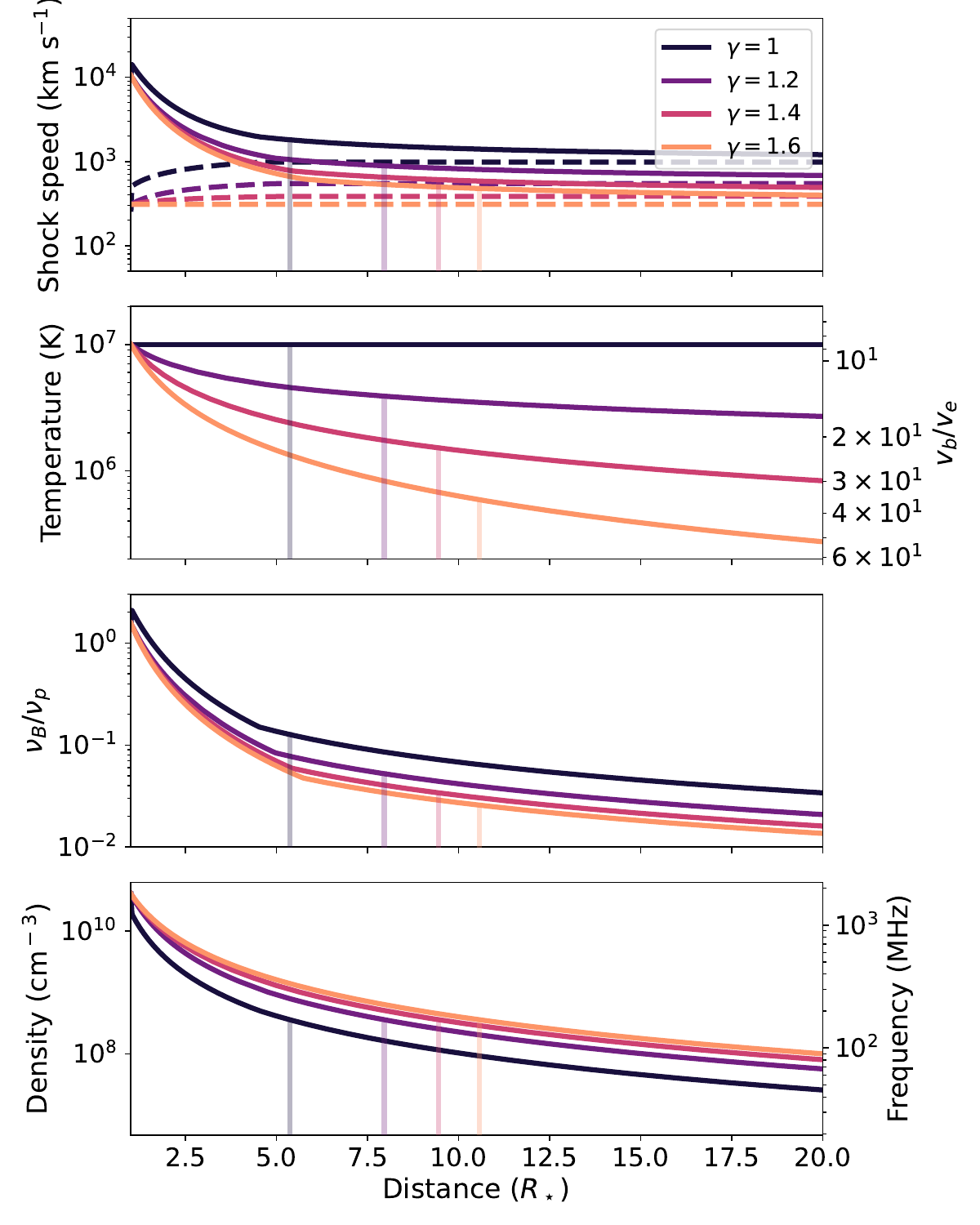}
    \caption{EK~Dra wind estimates for various polytropic indices. The coloured vertical lines indicate the distances corresponding to a frequency of 170\,MHz  ($h_{170}$). In the top plot, the dashed lines indicate the wind speed, and the solid lines show the super-Alfv\'enic shock requirements.}
    \label{fig:polytropic_profiles}
\end{figure}

It is physically impossible for the solar wind to be isothermal, and we indeed observe that the  $\sim10^6$\,K base temperature drops to $\sim10^{5}\,$K by $1\,$au \citep{Wilson2018, Mozer2023}.
Here, I demonstrate the effect of polytropic indices $1<\gamma<1.6$ on the wind of EK~Dra as it possesses the most extreme wind and burst deviations in these models. 
The wind profiles are shown in Fig.~\ref{fig:polytropic_profiles} and the relevant quantities are reported in 
Table~\ref{tab:polytropic_results}.
The general trend is that the higher the polytropic index, the slower the wind.
This makes sense. The decreasing temperatures leads to a decrease in the thermal pressure that accelerates the wind.
An effect of the reduced wind speed and pressure is that the Alfv\'en radius is pushed to farther distances as the kinetic energy associated with a given density is lower.
However, the smaller velocity gradient up to the Alfv\'en radius means that the density at a given distance is higher for larger indices.
In the case of $\gamma=1.6$, material needs to travel about twice as far to excite 170\,MHz plasma emission than in the isothermal case. 

\begin{figure}
    \centering
    \includegraphics[width=\linewidth]{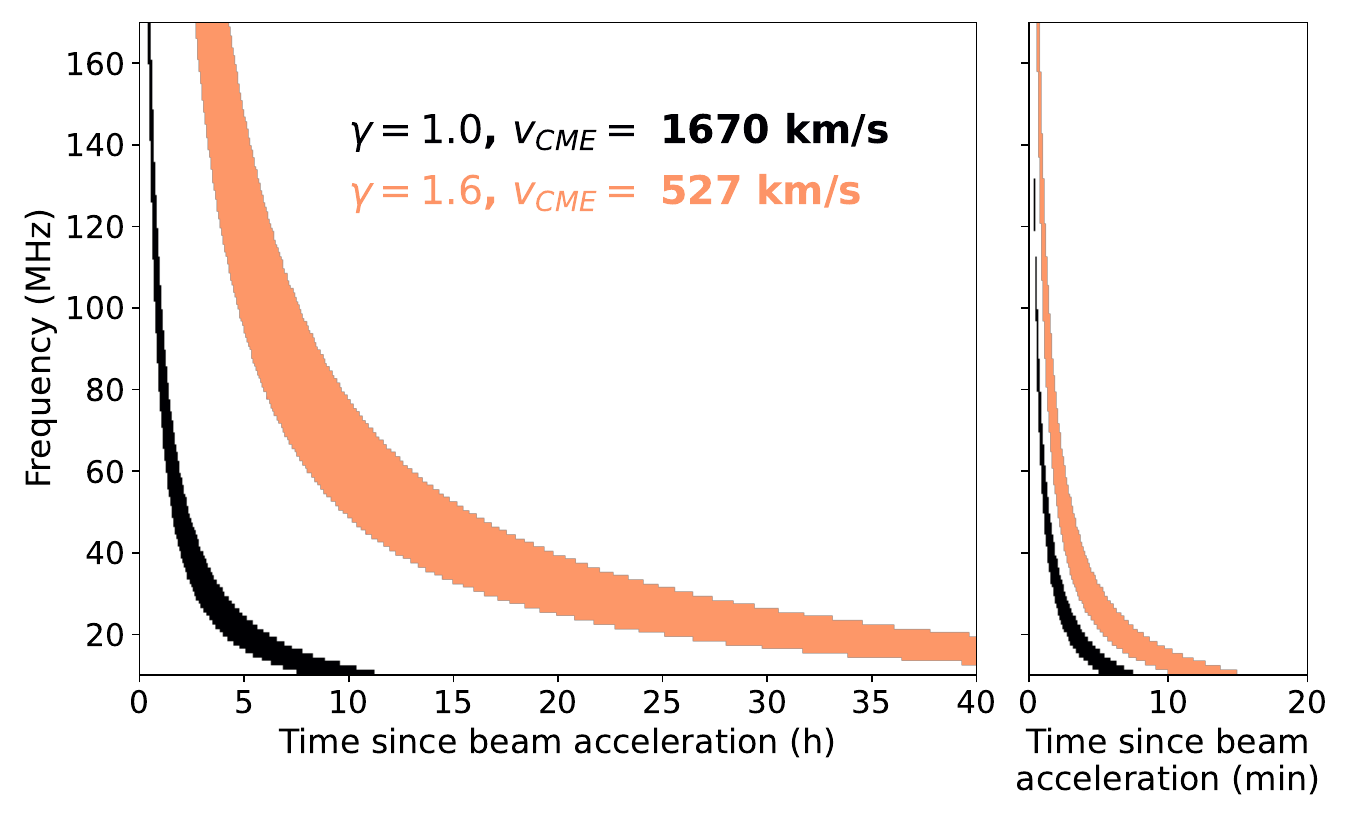}
    \caption{Dynamic spectra for type~II bursts (left) and type III bursts (right) for polytropic indices of 1.0 and 1.6.}
    \label{fig:polytropic_bursts}
\end{figure}

Thus, bursts that occur in winds with higher polytropic indices are delayed relative to bursts that occur in the isothermal wind.
I again modelled the type~II bursts as being $1.7\,v_\infty$ of the respective winds and the type~III bursts are for beams with speed $0.5\,c$. The burst profiles for $\gamma=1$ and 1.6 are shown in Fig.~\ref{fig:polytropic_bursts}.
It is worth noting the timescale of type~II bursts in the case of $\gamma=1.6$.
Even a shock-driving CME can require many hours to reach a distance conducive to exciting 170\,MHz plasma emission, and it may take days to excite $10\,$MHz. 
This long delay arises because the density gradient is much shallower than for $\gamma<1.6$ and because the shock requirements are so low. The speed required to drive a shock for $\gamma=1.6$ is about one-third of that for the isothermal scenario.
The burst expectations change when the CME moves faster, but for a CME or electron beam of a given speed, it will still take $\gtrsim2\times$ longer to excite 170\,MHz in the $\gamma=1.6$ case than in the isothermal scenario.
It is also worth noting that prominences observed from EK~Dra tend to have maximum line-of-sight velocities $\lesssim700\,$km\,s$^{-1}$ \citep{Namekata2025}. If we hope to observe a type~II burst from EK~Dra, we may need $\gamma >1$ for a shock to develop, and then, the burst itself may occur several hours after any H-$\alpha$ or other spectroscopic signature.

For the type~III bursts, a non-isothermal profile can make a rather significant impact on the beam velocity contrast.
Because the electron beam needs to travel so far to excite the relevant frequency, the thermal speed of the background plasma may drop significantly even for moderate values of $\gamma$. 
Such a change in the thermal population makes it straightforward for a beam that is only very mildly relativistic to achieve the $10\lesssim v_\text{b}/v_e\lesssim50$ range that is observed for solar type~III bursts.

\begin{table}[]
\caption{Key EK~Dra wind properties for various polytropic indices. }
    \centering
    \begin{tabular}{l c c c c c}
    \hline\hline
        {$\gamma$} &  & 1.0 & 1.2 & 1.4 & 1.6 \\\hline
        $r_\text{A}$ & $R_\star$ & 4.52 & 4.97 & 5.42 & 5.73 \\
        $r(\nu_B/\nu_p <1)$& $R_\star$ & 1.59 & 1.29 & 1.27 & 1.25 \\
         $h_{170}$ & $R_\star$ & 4.4 & 7.0 & 8.5 & 9.6 \\
         $v_\infty$& km\,s$^{-1}$ & 983 & 546 & 387 & 310\\
         $v_{s,170}$& km\,s$^{-1}$ & 1809 & 888 & 608 & 478 \\
         $T_{170}$& $10^6$\,K & 10.0& 3.9 & 1.5 & 0.6 \\
         $v_b/v_{e, 170}$& &  8.6 & 13.8 & 22.1 & 35.4 \\
         $\Delta t_{\text{II}, 170}$ & min & 16 & 45 & 76 & 107 \\
         $\Delta t_{\text{III}, 170}$ & s & 11 & 17 & 21 & 23 \\
         \hline
    \end{tabular}
    \tablefoot{The 170 subscript indicates that the property is evaluated at the distance at which the plasma density is conducive to 170\,MHz plasma emission. $\Delta t_\text{II, 170}$ and $\Delta t_\text{III, 170}$ refer to the duration of a type II and III burst, respectively.}
    \label{tab:polytropic_results}
\end{table}

\subsubsection{Alternative Alfv\'en distances}\label{sec:alternative_alf}
Because the Alfv\'en radius determines where the density and magnetic field gradients become less steep, it might have a meaningful impact on the arrival time of bursts and the relative contribution of the Alfv\'en speed to shock requirements.
Because the model here, and those employed by others \citep[e.g. ][]{Reville2016}, underestimates the Alfv\'en radius of the Sun, I assumed that the Alfv\'en radii for EK~Dra and AD~Leo predicted in this paper are likely also underestimates.

\begin{figure}
    \centering
    \includegraphics[width=\linewidth]{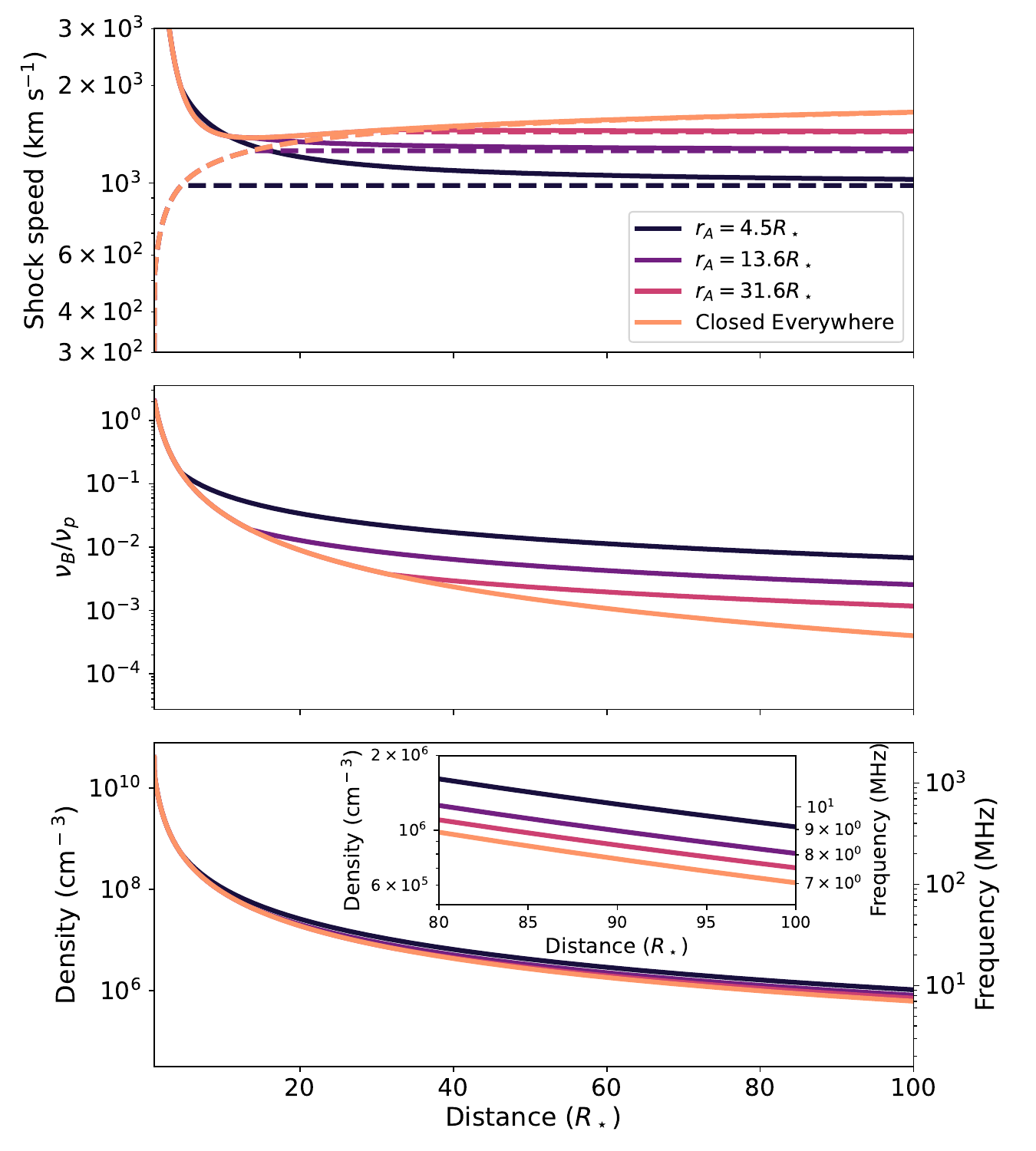}
    \caption{Isothermal wind solutions for various Alfv\'en distances. An inset for the number density in the 80$\,R_\star$--100$\,R_\star$ range is provided to better visualise differences between the resulting density profiles. In the top plot, the dashed lines indicate the wind speed, and the solid lines show the super-Alfv\'enic shock requirements.}
    \label{fig:varying_alf}
\end{figure}

I considered the cases where the Alfv\'en radius for EK~Dra is three times larger and seven times larger than the predicted Alfv\'en radius of {$4.5\,R_\star$} in the isothermal wind. 
I also considered the case that the field is closed everywhere.
The isothermal wind results for the different Alfv\'en distances are shown in Fig.~\ref{fig:varying_alf}.
As expected, larger Alfv\'en radii lead to higher wind speeds and to a lower relative contribution of the Alfv\'en speed to the shock requirements.
Thus, increasing the Alfv\'en radius makes it more difficult to achieve the necessary shock conditions to produce a type~II burst.
The overall impact on the density profile is minor, however. Although the producibility of a type~II burst changes substantially with Alfv\'en radius, the actual shape of a burst associated with a CME of a given speed does not change substantially with Alfv\'en radius.

\section{Summary and prospects}
Type~II and III bursts have long been established as potentially powerful tools for studying exo-solar space weather.
I demonstrated that the extreme environments of active stars can make such bursts more powerful than originally considered because they may probe distances far into the interplanetary medium. 
However, the strength of these bursts as space weather tools can also be limited by the exceptional environments through which the relevant accelerators propagate.
In particular, the high coronal temperatures produce strong winds and set highly energetic background electron populations.
Such winds and background populations make it difficult to produce the shocks and large velocity contrasts that are required for type~II and III bursts, respectively.

Additionally, the high densities that make low frequencies so effective for probing far distances also complicate coordinated observations when searching for type~II bursts; the most effective coordinated observation with other parts of the EM spectrum is unlikely to be simultaneous with the radio observation.
This is particularly true when the wind is not isothermal.
In this case, the shock and velocity contrast requirements are easier to achieve, but at the expense of delaying the arrival time of the burst to a given frequency.
In the case of type~II bursts, the delay can be on the order of days at the lowest frequencies accessible to ground-based observatories.
Although such a long delay can make simultaneous multi-wavelength observations ineffective for identifying bursts associated with flares, it might lead to opportunities in triggered radio observations.
In addition to the observational constraints, these results suggest that {we need to explore }a substantially different de-dispersion parameter space from the solar case {to identify} type~III bursts.

These results are born from a relatively simple one-dimensional framework.
While less rigorous than contemporary three-dimensional models, it replicates solar wind properties within the Alfv\'en radius and empirical relations for the duration and drift rate of a solar type~III burst reasonably well.
Although the wind predictions are likely less reliable beyond the Alfv\'en radius, the modelling framework presented here can be a reasonable first-order approximation for the burst structures of active stars.
When taken in tandem with the physical limitations of the plasma emission process, other coronal and wind properties such as the temperature profile might also be discerned, and at distances from the star that have historically been inaccessible.
Because the model is also computationally inexpensive, many parameters can be explored relatively quickly in order to provide rough estimates of stellar wind properties for stars producing candidate type II and III bursts. 

This work focused on AD~Leo and EK~Dra because these stars have been so well characterised and because they have candidate CME detections.
They were therefore valuable stars for demonstrating the wind and burst model predictions and to contextualise these predictions with respect to the mass motion observed at other wavelengths.
It is important to re-emphasise the growing number of candidate stellar plasma bursts \citep{Zic2020, Callingham2025, Wang2026}, however, including specifically from AD~Leo \citep{Mohan2024adleo}.
It is worth revisiting these bursts within the framework laid out here in order to better understand the coronal conditions under which these bursts may have developed.
By extension, the model I presented here can be used to more effectively distinguish between ECMI and plasma emission as the relevant emission process.

This work also focused on a relatively narrow frequency range compared to the $120\lesssim\nu\lesssim3000$\,MHz range that stellar bursts have been observed.
The $10<\nu<170$\,MHz range evaluated here was informed by instruments such as the OVRO-LWA and LOFAR, whose surveys are particularly well suited for stellar burst searches. It will be especially important to consider higher frequencies in the age of the Square Kilometre Array \citep[SKA, ][]{Dewdney2009}, however.
The low band of the SKA will cover 50--350\,MHz with an unprecedented sensitivity: on the order of $10\,$mJy at 1\,s and 1\,MHz time and spectral resolution, respectively \citep{Braun2019}.
This sensitivity is likely to substantially increase the population of stellar analogues to type~II and III bursts by extending the search space to the lower end of the burst luminosity function. 
Altogether, the recent increase in detections, the imminent improvement of low-frequency instruments, and the continued development of wind models and plasma theory are likely to substantially advance our understanding of stellar space weather in the near future.

\section*{Data availability}
The \texttt{swabs} code is available at \url{https://github.com/iveydavis/plasma\_analysis}.

\begin{acknowledgements}
I thank G.\,Hallinan and H.\,K. Vedantham for insightful discussions on plasma instability requirements, J.\,R. Callingham for additional discussions and comments on the manuscript{,  H.\,W. Edler for proofreading and testing the \texttt{swabs} code, and the anonymous referee for their helpful comments.}
\end{acknowledgements}

\bibliographystyle{aa} 
\bibliography{bib_file} 

\end{document}